\documentclass[nofootinbib,aps,prd,twocolumn,showpacs,superscriptaddress,footinbib]{revtex4-2} 
\usepackage{amssymb}
\usepackage{amsmath}
\usepackage[colorlinks=true,linkcolor=blue,citecolor=blue, urlcolor=blue]{hyperref} 
\usepackage{tikz}
\usepackage{graphicx}
\usepackage[utf8]{inputenc}
\definecolor{uibred}{RGB}{167, 38, 47}

\newcommand{\vp}{\mathbf{p}}
\newcommand{\vk}{\mathbf{k}}

\begin{document}

\title{Far from equilibrium hydrodynamics of nonthermal fixed points}

\author{J{\"u}rgen Berges} \email{berges@thphys.uni-heidelberg.de}
\affiliation{Institut f\"{u}r Theoretische Physik, Universit\"{a}t Heidelberg, 69120 Heidelberg, Germany}

\author{Gabriel S.\ Denicol} \email{gsdenicol@id.uff.br}
\affiliation{Instituto de F\'isica, Universidade Federal Fluminense, Niter\'oi, Rio de Janeiro, 24210-346, Brazil}

\author{Michal P.\ Heller} \email{michal.p.heller@ugent.be}
\affiliation{Department of Physics and Astronomy, Ghent University, 9000 Ghent, Belgium}

\author{Thimo Preis} \email{preis@thphys.uni-heidelberg.de}
\affiliation{Institut f\"{u}r Theoretische Physik, Universit\"{a}t Heidelberg, 69120 Heidelberg, Germany}

\begin{abstract}
Nonthermal fixed points are paradigmatic far-from-equilibrium phenomena of relevance to high-energy physics, cosmology, and cold atomic gases. We propose that, despite their intrinsically nonequilibrium nature, nonthermal fixed points give rise to hydrodynamic excitations otherwise known in the vicinity of thermal equilibrium. As a result, nonthermal fixed points can also be characterized by transport coefficients, such as a far-from-equilibrium, and therefore manifestly time-dependent, incarnation of shear viscosity. We corroborate our proposal with explicit studies using relativistic kinetic theory with binary collisions of massless particles in the 14-moment approximation and comparisons to QCD kinetic theory simulations.
\end{abstract}

\maketitle

\section{Introduction}

Our understanding of nonequilibrium physics is chiefly rooted in studying small departures from equilibria. This is in particular the origin of fluidity that gives rise to a wealth of phenomena across scales, from tiny droplets of quark-gluon plasma at RHIC and LHC, through cold atomic gases, many macroscopic phenomena on Earth, all the way to modeling astrophysical objects, and even to cosmological scenarios, see~\cite{Andersson:2006nr,Schafer:2009dj,Rezzolla:2013dea,Busza:2018rrf} for a broad exposition of these topics. Indeed, in its textbook form, fluid mechanics relies on a derivative expansion in gradients of fluid velocity and thermodynamic variables such as temperature or pressure around local thermal equilibrium~\cite{landau1959fm}.

In recent years, initially motivated by the physics of the quark-gluon plasma at RHIC and LHC~\cite{Heller:2015dha,Romatschke:2017vte} and now also by cold atomic gases~\cite{Fujii:2024yce,Mazeliauskas:2025jyi}, significant efforts have been invested into understanding far from equilibrium generalizations of fluid mechanics. This ongoing research program, reviewed in~\cite{Soloviev:2021lhs,Jankowski:2023fdz}, has been concerned with particular resummations of hydrodynamic derivative expansion with the aim of extending its validity deeper into a nonequilibrium regime.

In the present work, inspired by these ongoing efforts to push hydrodynamics far from equilibrium, we propose that fluid mechanics naturally emerges as an effective description of the evolution of spatial inhomogeneities around nonthermal fixed points. Nonthermal fixed points are intrinsically far from equilibrium phenomena, whose defining feature is self-similarity of the distribution of particles in time~\cite{Micha:2002ey,Berges:2008wm,Berges:2013eia,Kurkela:2015qoa}. They are of broad theoretical relevance from cosmology, through thermalization in QCD all the way to cold atomic gases, see~\cite{Berges:2020fwq,Mikheev:2023juq} for reviews. Furthermore, in the latter case, nonthermal fixed points have been realized experimentally in several complementary cold atomic setups~\cite{Prufer:2018hto,Erne:2018gmz,Glidden:2020qmu,
Navon:2016em,johnstone2019evolution,Helmrich:2020sgn,Garcia-Orozco:2021hkx,Martirosyan:2023mml,Huh:2023xso,Lannig:2023fzf,Gazo:2023exc,Martirosyan:2024rxm}. This adds a layer of observational relevance to the novel manifestation of fluidity that we propose.

Our proposal that nonthermal fixed points support fluid mechanics should come across as a conservative one. One way to motivate standard, near-equilibrium hydrodynamics, at least at a linearized level, is the attractive nature of equilibrium combined with the presence of conservation laws. Local excesses of energy, momentum, or other conserved charges cannot dissipate arbitrarily quickly as their dynamics are constrained by the microscopic conservation law. Instead, equilibrium is approached via the transport of conserved quantities across the system, which is the defining feature of hydrodynamics. Similar assumptions hold around (at least) isotropic nonthermal fixed points: ab initio simulations, experimental observations, and recent results on the spectrum of linearized perturbations~\cite{Preis:2022uqs,DeLescluze:2025jqx} indicate their stability and local conservation laws hold.

In practical terms, nonthermal fixed points studied to date were (on average) spatially homogeneous. The core of our idea is that spatial inhomogeneity around a nonthermal fixed point relaxes according to hydrodynamic equations of motion with transport coefficients, the most important of which is the shear viscosity, being intrinsically time-dependent. Note that ultimately this notion of far from equilibrium hydrodynamics is novel with respect to the efforts that motivated us. The reason for it is that already our point of departure for constructing a hydrodynamic expansion is far from equilibrium and this is ultimately the source of time-dependent transport coefficients, rather than a resummation of the near-equilibrium derivative expansion. 

Our work naturally connects with two major research directions of the past two decades: studying patterns of real-time response of quantum field theories near equilibrium in different parametric regimes~\cite{Kovtun:2005ev,Hartnoll:2005ju,Romatschke:2015gic,Grozdanov:2016vgg,Kurkela:2017xis,Moore:2018mma,Ochsenfeld:2023wxz,Brants:2024wrx} and, related to it, comparing shear viscosity of different physical systems that led to consideration of its ratio with entropy density~\cite{Kovtun:2004de}. The novel aspect of our results in this context stems from the starting point of our analysis, the underlying nonthermal fixed point, being intrinsically time dependent and far from equilibrium. In particular, one resulting problem that we address is how to compare time-dependent shear viscosity between nonthermal fixed points or with a shear viscosity of the same underlying system near equilibrium, given that its ratio with statistical entropy turns out to retain time dependence. The latter obstructs the usage of the shear viscosity to entropy density ratio in this context.

While our idea about far from equilibrium fluidity around nonthermal fixed points is general, the realization we pursue is in kinetic theory for binary collisions of particles and comparisons of our predictions with QCD kinetic theory simulations~\cite{Arnold:2002zm,Arnold:2003zc,Kurkela:2015qoa,Mazeliauskas:2018yef}. In our studies, we will rely on the 14-moment approximation that has previously been used to construct near-equilibrium hydrodynamic equations of motion from kinetic theory, see~\cite{Denicol:482569,Rocha:2023ilf} for a review, and Ref.~\cite{Denicol:2021wod} for an initial discussion of far from equilibrium hydrodynamics in kinetic theory.

\vspace{12 pt}

\noindent \textbf{Note about conventions and notation:} In our paper, we use natural units $k_{B} = \hbar = c = 1$ and adopt the mostly minus metric signature convention.  We mostly specialize to near-isotropic scale-invariant (massless) systems in three spatial dimensions, for which the background energy density $\cal E$ and pressure $\cal P$ are related as ${\cal E} = \, 3\, {\cal P}$ and the bulk viscosity vanishes. Further notation is specified in Appendix \ref{Appendix:Notation}.

\section{Hydrodynamics of nonthermal fixed points}

\subsection{Universal scaling far from equilibrium}
We will consider perturbations of isotropic (in the local rest frame of the system) nonthermal fixed points, where correlation functions exhibit characteristic self-similar behavior in a corresponding momentum regime
\begin{equation}
\label{eq.prescaling}
    f(t,\vk) = A(t) f_S(B(t)|\vk|).
\end{equation}
In the equation above, $f(t,\vk)$ is the one particle distribution function subject to the spatially homogeneous Boltzmann equation. The factors $A(t)$ and $B(t)$ are power laws in time
\begin{equation}
    A(t) = \left(\frac{t-t_*}{t_{\mathrm{ref}}} \right)^{\alpha},\quad B(t) = \left(\frac{t-t_*}{t_{\mathrm{ref}}} \right)^{\beta},
\end{equation}
whose exponents together with the scaling function $f_S$ constitute universal scaling properties far from equilibrium. The time offset~$t_{*}$ is needed for the system to dynamically account for the notion of the origin of time, which plays a special role for scaling phenomena~\cite{Heller:2023mah,Gazo:2023exc}.

Through the lenses of Eq.~\eqref{eq.prescaling}, nonthermal fixed points are self-similar turbulent cascades transporting conserved quantities in the given momentum range. We will focus on nonthermal attractors characterized by local mode energy conservation, i.e., where $\mathcal{E}=\int d^d\vk/(2\pi)^d \omega_\vk f(t,\vk)$ is constant in the momentum range. Using Eq.~\eqref{eq.prescaling}, this implies the relations \begin{equation}
    \sigma\equiv \log(A(t))/\log(B(t))=d+z=\alpha/\beta,
\end{equation}
where~$d$ is the dimensionality of space (for us, $d = 3$) and $z$ the dynamical scaling exponent that can be deduced from the underlying (quasi-)particle dispersion relation $\omega_\vk\sim |\vk|^z$ (for us, $z = 1$).

\subsection{Hydrodynamics}

The main equations governing the dynamics of a relativistic fluid emerge from the conservation laws for energy and momentum \footnote{In this work, we neglect the dynamics of any conserved charge.},
\begin{align}
\label{eq.hydro_macro}
    \partial_\mu T^{\mu\nu} &= 0,
\end{align}
with $T^{\mu\nu}$ being the energy-momentum tensor of the fluid. It is convenient to decompose $T^{\mu\nu}$ in terms of the fluid's four-velocity, $u^\mu$, 
\begin{equation}
T^{\mu \nu} = \mathcal{E} u^\mu u^\nu - \mathcal{P} \Delta^{\mu \nu} + \pi^{\mu \nu},
\end{equation} 
where $\mathcal{E}$ is the energy density in the local rest frame of the fluid, $\mathcal{P}$ is the thermodynamic pressure, and $\pi^{\mu \nu}$ is the shear stress tensor. The focus of this paper will be conformal systems, leading to the relation $\mathcal{P}=\mathcal{E}/3$ regardless of whether the system is close to equilibrium. We note that in the tensor decomposition above, $u^\mu$ is defined according to the Landau picture~\cite{landau1959fluid} as the unit-normalized ($u_\mu u^\mu =1$) timelike eigenvector of the energy-momentum tensor, satisfying $T^\mu_\nu u^\nu = \mathcal{E} u^\mu$.

Naturally, the continuity equations describing energy-momentum conservation are not sufficient to completely determine the dynamics of the energy-momentum tensor -- which contains a total of 9 independent degrees of freedom since $\mathcal{E}=3\mathcal{P}$. In order to close this system of equations one must provide additional constraints for the shear stress tensor. In the relativistic regime, causality and linear stability around global equilibrium dictate that the shear stress tensor cannot be simply determined by constitutive relations, as is often the case in the non-relativistic regime \cite{landau1959fluid}, satisfying instead dynamical equations of motion. The most widespread theory of relativistic hydrodynamics was proposed by Israel and Stewart in the 1970's \cite{israel1979transient}. This formulation of relativistic hydrodynamics is widely employed in simulations of ultrarelativistic heavy ion collisions to describe the time evolution of the quark-gluon plasma \cite{Gale:2013da,Paquet:2023rfd}. It has also been systematically tested in the domain of kinetic theory \cite{Denicol:482569}.

In the Israel and Stewart approach, the shear stress tensor $\pi^{\mu \nu}$ obeys relaxation-type equations of motion
\begin{align}
\label{eq.eom_shear_thermal}
    \tau_\pi \dot{\pi}^{\langle \mu \nu\rangle}  &=-\pi^{\mu \nu}+2\eta \sigma^{\mu \nu} + \ldots\,,
\end{align}
with the shear tensor $\sigma^{\mu \nu}$ defined by
\begin{equation}
\label{sigmadef}
\sigma^{\mu \nu}\equiv \partial^{\langle \mu} u^{\nu \rangle},
\end{equation}
and the ellipsis denoting possible nonlinear terms. The brackets denote the double, symmetric, and traceless projection of the tensor, as defined in Appendix \ref{Appendix:Notation}. In Eq.~\eqref{eq.eom_shear_thermal}, $\eta$ is the shear viscosity, which also appears in Navier-Stokes theory, and $\tau_\pi$ is the relaxation time, a novel parameter that appears in relativistic hydrodynamic equations of motion and quantifies the time it takes for the shear stress tensor to relax back to equilibrium. Linear causality and stability around equilibrium for massless particles (conformal system) dictate that it has a lower bound~\cite{Pu:2009fj}
\begin{equation}
\label{eq.causalityineq}
\frac{\eta}{\tau_{\pi} \left({\cal E}+{\cal P}\right)}
 \leq \frac{1}{2}.
\end{equation}
The combination appearing on the left-hand side of the above inequality will play an important role later in the paper.

The equations of motion for relativistic hydrodynamics of the type~\eqref{eq.eom_shear_thermal} can be derived from the Boltzmann equation~\cite{deGroot:80relativistic}, which
provides a microscopic description of dilute gases in terms of the single
particle momentum distribution function, $f(x,k) \equiv f_{\mathbf{k}}$. The Boltzmann equation is
a partial integro-differential equation that determines the time evolution
of $f_{\mathbf{k}}$, 
\begin{equation}
k^{\mu }\partial _{\mu }f_{\mathbf{k}}=\mathcal{C}[f],
\end{equation}
with $k^{\mu }=(\omega _{\mathbf{k}},\mathbf{k})$ being the on-shell
four-momentum and $\mathcal{C}[f]$ the collision kernel encoding microscopic interactions. Here we consider a collision term with elastic two-to-two collisions in the overoccupied limit, $f_{\mathbf{k}}\gg 1$,
\begin{align}
C\left[ f\right] & \approx \frac{1}{2}\int dK^{\prime }dPdP^{\prime }W_{\mathbf{kk}\prime \rightarrow \mathbf{pp}\prime } \notag \\
& \times \left[(f_{\mathbf{p}}f_{\mathbf{p}^{\prime }} 
 \left( f_{\mathbf{k}}+f_{%
\mathbf{k}^{\prime }}\right) -f_{\mathbf{k}}f_{\mathbf{k}^{\prime }}\left(
f_{\mathbf{p}}+f_{\mathbf{p}^{\prime }}\right)\right],  \label{Col_term}
\end{align}%
where $W_{\mathbf{kk}\prime \rightarrow \mathbf{pp}\prime }$ is the
Lorentz-invariant transition rate.
We will focus on nonthermal attractors associated with the
turbulent transport of mode energy to smaller distance scales through a
direct energy cascade. The underlying physics is captured using
weakly-interacting relativistic massless scalars, where the corresponding
transition element for example in the $O(N)$ model reads~\cite%
{PineiroOrioli:2015cpb} 
\begin{align}
W_{\mathbf{kk}\prime \rightarrow \mathbf{pp}\prime }& =\lambda ^{2}\frac{N+2%
}{48N^{2}}  \notag  \label{eq.transition_matrix} \\
& \times \left( 2\pi \right) ^{d+1}\delta ^{\left( d+1\right) }\left( k^{\mu
}+k^{\prime \mu }-p^{\mu }-p^{\prime \mu }\right),
\end{align}%
with $\omega _{\mathbf{k}}=|\mathbf{k}|$.

By
taking appropriate moments of this equation, one recovers the conservation
laws in Eq.~\eqref{eq.hydro_macro} and identifies the following microscopic
expression for $T^{\mu \nu }$ 
\begin{equation}
T^{\mu \nu }=\int dKk^{\mu }k^{\nu }f_{\mathbf{k}},
\end{equation}%
where $dK=d^{3}\mathbf{k}/[(2\pi )^{3}\omega _{\mathbf{k}}]$ is the
Lorentz-invariant phase volume. The local energy density and the
shear stress tensor can then be identified as the following moments~of~$
f_\vk$,
\begin{subequations}
\begin{eqnarray}
\mathcal{E} &=&\int dK\left( u_{\mu }k^{\mu }\right) ^{2}f_{\mathbf{k}}, \\
\pi^{\mu \nu } &=&\int dKk^{\langle \mu }k^{\nu \rangle }f_{\mathbf{k}}.\label{eq.pimunukin}
\end{eqnarray}
\end{subequations}
We note that $\pi^{\mu \nu }$ vanishes when integrated over any isotropic momentum distribution and, hence, will also vanish when integrated over the isotropic nonthermal fixed point distribution of Eq.~\eqref{eq.prescaling}. Perturbations around the nonthermal fixed point can then lead to finite values of the shear stress tensor that, as we shall demonstrate, can be described using hydrodynamic-like equations. 

The fundamental step in the derivation of a fluid-dynamical 
theory is the expansion of the distribution function about some background state, 
\begin{equation}
f_\vk=f_{0\vk}+\delta f_\vk,  \label{eq.lr_bg_plus_pert}
\end{equation}%
and then deriving an approximate expression for the correction $\delta f_\vk
$. Since the fluid-dynamical regime is traditionally expected to emerge if
the system is sufficiently close to equilibrium, $f_{0\vk}$ is usually assumed
to be the local equilibrium distribution function and $\delta f_\vk$ the corresponding nonequilibrium correction. One then applies the 14-moment approximation \cite{israel1979transient} and expresses the correction $\delta f_\vk$ solely in terms of the shear stress tensor (and other moments of $f_\vk$, when applicable), effectively closing the hydrodynamic equations of motion. In this work, we extend this procedure and take $f_{0\vk}$ to be an isotropic nonthermal fixed point distribution as in Eq.~\eqref{eq.prescaling}. We do so by noting that the traditional 14-moment approximation does not necessarily rely on the assumption that $f_{0\vk}$ is the local equilibrium distribution: it can be directly applied to expansions around any \textit{isotropic} momentum distribution function. One then obtains the direct extension of the 14-moment approximation \cite{Denicol:2021wod},
\begin{equation}
\label{eq.phi_14moments}
    \delta f_\vk =  f_{0\vk} \frac{2p^\mu p^\nu}{15 N_4(t)} \pi_{\mu \nu}(t),
\end{equation}
with 
\begin{equation}
N_4(t)=\int dK |\vk|^4 f_{0\vk}   , 
\end{equation} 
and $f_{0\vk}$ being the isotropic nonthermal fixed point distribution from Eq.~\eqref{eq.prescaling}. More details can be found in Appendix \ref{Appendix:hydro}.

Next, following the procedure outlined in~\cite{Denicol:2010xn,Denicol:2021wod},
we obtain an equation of motion for $\pi ^{\mu \nu }$ by taking the
comoving derivative of its kinetic expression~\eqref{eq.pimunukin}, 
\begin{equation}
\dot{\pi}^{\langle \mu \nu \rangle }=\Delta _{\alpha \beta }^{\mu \nu }\frac{d}{d\tau }%
\int dKk^{\langle \alpha }k^{\beta \rangle }f_\vk,
\end{equation}%
and using the Boltzmann equation to replace the comoving derivatives of $f_\vk$, together with the 14-moment approximation \eqref{eq.phi_14moments}.
Considering this more general application of the method of moments, one can
generalize the derivation of fluid dynamics proposed by Israel and Stewart
to a broader class of systems that are not necessarily close to equilibrium. In this work, we perform this task for systems that are in the vicinity of a nonthermal attractor, leading to a
fluid-dynamical description of the fluctuations around this scaling background.

The evolution equation of the shear stress becomes,
\begin{align}
\label{eq.eom_shear}
    \tau_\pi(t) \dot{\pi}^{\langle \mu \nu\rangle} &=- \pi^{\mu \nu} +2\eta(t) \sigma^{\mu \nu}\\
    &+\tau_\pi(t) \left[-\frac{4}{3}\pi^{\mu \nu}\theta-\frac{10}{7} \pi^{\lambda \langle \mu}\sigma^{\nu\rangle}_\lambda -2\pi^{\lambda\langle \mu} \omega^{\nu\rangle}_\lambda \right]\nonumber,
\end{align}
where the expansion rate $\theta$, the vorticity tensor $\omega^{\alpha \beta}$, and details of this derivation are given in Appendix \ref{Appendix:hydro}. In direct analogy with the thermal case \eqref{eq.eom_shear_thermal}, 
we identified the shear viscosity and relaxation time transport coefficients as
\begin{subequations}
\label{eq.transport_coeff}
\begin{align}
    \eta(t) &= \frac{8}{15^2} \frac{\mathcal{E} N_4(t)}{\delta \mathcal{C}(t)},\\
        \tau_\pi(t) &= \frac{2}{15}\frac{N_4(t)}{\delta \mathcal{C}(t)}.
\end{align}
\end{subequations}
The quantity $\delta \mathcal{C}(t)$ emerges from the linearized collision kernel and encodes all information about the microscopic interactions; an exact expression is given in Eq.~\eqref{eq.coll_linearized} of Appendix \ref{Appendix:hydro}. Crucially, in comparison with its counterpart~\eqref{eq.eom_shear_thermal} derived in an expansion around local thermal equilibrium, the transport coefficients following from an expansion around the scaling background display an intrinsic time dependence. We note that Eq.~\eqref{eq.eom_shear} also contains the nonlinear terms omitted from~\eqref{eq.eom_shear_thermal}, as well as their corresponding transport coefficients.

\subsection{Comparisons with equilibrium}

Note that, without dwelling into details of Eq.~\eqref{eq.transport_coeff}, we can already conclude that the ratio of the shear viscosity to the relaxation time is time-independent far from equilibrium. This is important, as it allows for making a comparison with equilibrium hydrodynamics across microscopic models. A powerful notion considered before near equilibrium is the dimensionless ratio of the shear viscosity to the entropy density~\cite{Kovtun:2004de}. Its importance stems from connecting a very basic hydrodynamic notion with the microscopic interaction strength in a way that allows for a straightforward distinction between small and large couplings. However, as one can convince oneself, the $\eta/s$ ratio evaluated on a nonthermal fixed point exhibits an unilluminating time dependence, which renders it useless, at least for near-equilibrium comparisons.

What eliminates the time dependence is precisely the ratio $\eta/\tau_\pi$. Its drawback, however, is that it is dimensionful. To make it dimensionless, we can use the energy density or pressure, but it is more insightful to consider the combination of ${\cal E} + {\cal P}$, as we did in~\eqref{eq.causalityineq}. The resulting quantity is
\begin{equation}
\label{eq.etacombination}
\frac{\eta(t)}{\tau_{\pi}(t) \left({\cal E}+{\cal P}\right)} = \frac{1}{5}.
\end{equation}
Since the derivation behind~\eqref{eq.transport_coeff} relied only on the isotropy of $f_{0\vk}$, the expression~\eqref{eq.etacombination} applies also to the standard near-equilibrium hydrodynamics. This indicates a degree of universality between the equilibrium and nonthermal fixed point hydrodynamics for the same underlying weakly-coupled quantum field theory.

In holography at infinite coupling, the shear viscosity is universally given by $\eta = \frac{1}{4\pi} \frac{{\cal E} + {\cal P}}{T}$, where we used thermodynamic identities to rephrase the entropy density in terms of the variables used in~\eqref{eq.etacombination}. On the other hand, the relaxation time is simply the gap in the spectrum of quasinormal modes associated with the energy-momentum tensor sector~\cite{Kovtun:2005ev}. This gap changes once the conformal symmetry is relaxed~\cite{Buchel:2015saa,Janik:2015waa}, but in the absence of chemical potential these changes are of limited capacity when expressed in the units of the temperature. From the results of~\cite{Kovtun:2005ev}, we take $\tau_{\pi} T \approx 1.3\times 2\pi$. This leads to
\begin{equation}
\frac{\eta}{\tau_{\pi} \left({\cal E}+{\cal P}\right)} \approx 0.65.
\end{equation}
Interestingly, this is the same order of magnitude as the weak coupling result~\eqref{eq.etacombination} and goes hand-in-hand with earlier attempts to compare weak and strong coupling dynamics by normalizing time by $\eta/s$ and making it dimensionless using temperature, as in~\cite{Heller:2016rtz,Du:2022bel}. This is in stark contrast with $\eta/s$, which by construction is parametrically different at weak and strong coupling.

\subsection{Scaling analysis} 

The kinetic theory origin of nonthermal fixed points can be traced to being able to factor out the time dependence from the collision kernel when evaluated on the overoccupied distribution function of the scaling form~\eqref{eq.prescaling}~\cite{Heller:2023mah}
\begin{equation}
    \mathcal{C}[f](t,\vp) = B^{1-1/\beta}(t) \mathcal{C}[f_S](B(t)|\vp|).
\end{equation}
Using a perturbative transition matrix element $W\sim \delta^{(d+1)}(k^\mu+{k^\prime}^\mu -p^\mu -{p^\prime}^\mu)$ like in~\eqref{eq.transition_matrix}, one can perform a general scaling analysis of the time-dependent transport coefficients
\begin{equation}
    \eta(t) = B^{4-4z+1/\beta}(t) \bar{\eta},
\end{equation}
with $\tau_{\pi}(t)$ scaling in the same way as dictated by Eq.~\eqref{eq.etacombination}. To make the notation uniform, we will write
\begin{equation}
    \tau_{\pi}(t) = B^{4-4z+1/\beta}(t) \bar{\tau}_{\pi}.
\end{equation}
The dimensionful constants $\bar{\eta}$ and, via Eq.~\eqref{eq.etacombination}, ${\bar{\tau}_{\pi}} = \frac{15}{4}\bar{\eta}/{\cal E}$  arise from evaluating Eqs.~\eqref{eq.transport_coeff} on the scaling function $f_{S}$. In the following, we will be primarily interested in the scaling properties associated with the factors of~$B^{4-4z+1/\beta}(t)$ and their implications for the dynamics, rather than in the precise values of $\bar{\eta}$ and $\bar{\tau_{\pi}}$.

\section{The relativistic direct energy cascade}
\label{sec.relativistic}

We will now focus on the dynamics of hydrodynamic perturbations around the nonthermal energy attractor in relativistic systems, which is of importance to early-universe cosmology~\cite{Micha:2002ey,Micha:2004bv,Berges:2008wm} and quark-gluon plasma physics~\cite{Kurkela:2012hp,AbraaoYork:2014hbk,Schlichting:2012es,Heller:2023mah}.

Starting from a representative of the corresponding universality class, such as e.g. relativistic weakly-coupled scalars with transition matrix element as in Eq.~\eqref{eq.transition_matrix}, the analysis of the Boltzmann equation leads to the following scaling exponents
\begin{equation} 
\label{eq.expts_UVrel}
\alpha = -\frac{4}{7} \quad \mathrm{and}\quad \beta = -\frac{1}{7},
\end{equation}
which were originally introduced in Eq.~\eqref{eq.prescaling}.
We note that the same scaling exponents arise for the QCD effective kinetic description~\cite{Schlichting:2012es,Berges:2013fga,AbraaoYork:2014hbk,Heller:2023mah}.

Utilizing the values~\eqref{eq.expts_UVrel}, we determine the underlying time dependence of the transport coefficients,
\begin{subequations}
\label{eq.scaling_qcd}
\begin{align}
    \eta(t) &= B^{-7}(t)\bar{\eta} = \left( \frac{t-t_{*}}{t_{\mathrm{ref}}}\right)\bar{\eta},\\
    \tau_\pi(t) &=  B^{-7}(t) \bar{\tau}_\pi= \left( \frac{t-t_*}{t_{\mathrm{ref}}}\right)\bar{\tau}_\pi.
\end{align}
\end{subequations}
This implies that both the shear viscosity and the relaxation time grow linearly with time. This behavior has an important consequence for the decay to a nonthermal fixed point, which we will explore next.

\subsection{Homogeneous isotropization dynamics \label{sec.isotropization}}

The far-from-equilibrium hydrodynamic equations derived in the previous section from kinetic theory~\eqref{eq.eom_shear} have a particularly simple sector associated with spatial homogeneity, in which case they reduce to
\begin{equation}
\label{eq.isotropization}
    \tau_\pi(t) \dot{\pi}^{\langle \mu \nu \rangle} + \pi^{\mu \nu}=0.
\end{equation}
Let us emphasize that this equation does not describe hydrodynamic excitations, but instead is a consequence of the natural framework we applied that approximates kinetic theory in a way encapsulating leading hydrodynamic effects.

Building on the earlier studies in~\cite{Chesler:2008hg,Heller:2012km,Heller:2013oxa,Chesler:2013lia}, we adopt the ansatz
\begin{equation}
\label{Ansatz_Pi}
    \pi^{\mu \nu} = \mathrm{diag}\left(0,\frac{1}{3} \Delta {\cal P}(t) , \frac{1}{3} \Delta {\cal P}(t),-\frac{2}{3} \Delta {\cal P}(t) \right),
\end{equation}
where $\Delta {\cal P}(t)$ characterizes the time-dependent pressure anisotropy. Eq.~\eqref{eq.isotropization} is then solved by
\begin{align}
\label{eq.deltaPsol}
        \Delta {\cal P}(t) &= \Delta {\cal P}(t_0) \exp\left[ -\int_{t_0}^t \frac{dt^\prime}{\tau_\pi(t^\prime)}\right]\nonumber\\
        &=\Delta {\cal P}(t_0) \left(\frac{t_0-t_*}{t-t_*} \right)^{\frac{t_{\mathrm{ref}}}{\bar{\tau}_\pi}},
\end{align}
 where, in the last step, we used the relaxation time calculated assuming an energy cascade from Eq.~\eqref{eq.scaling_qcd}. This is to be compared with the thermal case in which $\tau_\pi$ is simply a constant determined from the background temperature and the pressure anisotropy relaxes exponentially to zero,
 \begin{equation}
     \Delta {\cal P}(t) \Big |_{\mathrm{thermal}} = \Delta {\cal P}(t_0) \exp\left( - \frac{t-t_0}{\tau_\pi}\right).
 \end{equation}
 Similar exponential relaxation is seen in various causal formulations of relativistic hydrodynamics and in holography~\cite{Florkowski:2017olj}. Indeed, the reason behind the sub-exponential isotropization around the nonthermal fixed point is the effective relaxation time that increases with time. Note that this is in line with recent results~\cite{DeLescluze:2025jqx} about \emph{isotropic} relaxation to the same non-thermal fixed point going beyond Eq.~\eqref{eq.eom_shear}, which also exhibits a power-law rather than exponential approach. Finally, let us emphasize that Eq.~\eqref{eq.eom_shear} is an effective approximation to the full kinetic theory dynamics and we do not stipulate that the kinetic theory in question does have precisely an excitation described by~\eqref{eq.deltaPsol}. Indeed, we expect the nonhydrodynamic sector of kinetic theory to be much richer based on earlier near-equilibrium~\cite{Romatschke:2015gic,Kurkela:2017xis,Moore:2018mma,Ochsenfeld:2023wxz,Brants:2024wrx} and nonthermal fixed point studies~\cite{DeLescluze:2025jqx}.

The result~\eqref{eq.deltaPsol} will serve as a basis for a qualitative comparison with the QCD kinetic theory in Sec.~\ref{sec.qcd}.

\subsection{Hydrodynamic excitations and linear response theory}
\label{subsec.modes_rel}
At a linearized level around equilibrium, hydrodynamic excitations manifest themselves as gapless eigenmodes of the equilibrium counterpart of equations like Eq.~\eqref{eq.eom_shear}. Both equilibrium and known nonthermal fixed points are homogeneous, as a result of which it is natural to seek for perturbations in a Fourier space. Due to isotropy, we can align the spatial momentum with any of the directions, e.g. with $x^{3}$: $e^{i \, q\, x^{3}}$. Note that $q$ is the spatial momentum of perturbations associated with macroscopic inhomogeneity in space, not to be confused with the momentum of microscopic particles.

Now, in equilibrium it is natural to consider the time-dependence in the standard Fourier form, $e^{-i \Omega(q) (t-t_{0})}$. However, since both the shear viscosity and the relaxation time are time-dependent, we will instead consider a more general expression, $e^{-i \int_{t_0}^{t}dt' \Omega(t',q)}$, which in the absence of time dependence in $\Omega(t,q)$ reduces to the usual Fourier analyses performed for equilibrium states. The statement of gaplessness of hydrodynamics is simply $\Omega(t,q) \rightarrow 0$ as $q\rightarrow 0$, i.e. perturbations of arbitrarily small spatial momentum become arbitrarily long-lived and slowly evolving. The origin of this lies in the conservation equations, for us Eq.~\eqref{eq.hydro_macro}, in which all terms contain first-order derivatives.

There is one more standard technicality to take into account. The perturbations we consider are tensor perturbations and the choice of the spatial momentum gives rise to a residual $SO(2)$ symmetry. As a result, different irreducible representations of this $SO(2)$ decouple in their dynamics, giving rise to sound wave perturbations and shear modes, see e.g.~\cite{Kovtun:2005ev,Baier:2007ix} for discussions.

Starting with the unperturbed nonthermal fixed point having $\cal E$ and $u^{0} =1$, $u^{1,2,3} = 0$, $\pi^{\mu \nu} = 0$, the sound channel corresponds to $\delta {\cal E}$, $\delta u^{3}$, $\delta \pi^{33}$ and $\delta \pi^{11} = \delta \pi^{22} = -\frac{1}{2} \delta \pi^{33}$. The shear channel is then encapsulated by $\delta u^{1}$ and $\delta \pi^{13}$ with other possible components obtainable by a rotation around the $x^{3}$ axis.

It turns out that, after defining
\begin{subequations}
\label{eq.rescalingsomegaq}
\begin{eqnarray}
\bar{\Omega}(\bar{q}) &=&   \Omega(t,q) \times \left(\frac{t-t_{*}}{t_{ref}}\right),\\
\bar{q} &=& q \times \left(\frac{t-t_{*}}{t_{ref}}\right),
\end{eqnarray}
\end{subequations}
the dispersion relations $\bar{\Omega}(\bar{q})$ become not only time-independent, but also become identical to their near-equilibrium counterparts~\cite{Baier:2007ix}. This implies that at least some of the lessons learned from these equations will generalize to our far from equilibrium system. 

For the shear mode, the dispersion relation solves
\begin{equation}
\label{eq.shearmode}
\bar{\Omega}(\bar{q})+i \frac{3 \bar{\eta}}{4 {\cal E}} \bar{q}^2 - i \bar{\tau}_{\pi} \bar{\Omega}(\bar{q})^2 = 0.
\end{equation}
There are two solutions of this equation. The first one is hydrodynamic and describes diffusion at low $q$
\begin{equation}
\bar{\Omega}_{\mathrm{shear}}(\bar{q}) \approx - i \frac{3\bar{\eta}}{4{\cal E}} \bar{q}^2 ,
\end{equation}
with the combination $\frac{3\bar{\eta}}{4{\cal E}}$ playing the role of the diffusivity constant. Note that the quantity that ultimately dictates the time-dependence is $e^{-i \int_{t_0}^{t}dt' \Omega(t',q)}$, which means that at low-$q$ the shear diffusion far from equilibrium will proceed at late times as $e^{- \frac{3\bar{\eta}}{4{\cal E}}q^2 \, \left( \frac{t-t_{*}}{t_{\mathrm{ref}}} \right)^2}$, whereas near-equilibrium the behavior is~$e^{- \frac{3\eta}{4{\cal E}}q^2 \, t}$. Despite this qualitative change of the behavior with respect to near-equilibrium dynamics, the far from equilibrium diffusive mode still gets arbitrarily long-lived as $q \rightarrow 0$. The other excitation associated with~\eqref{eq.shearmode} should be viewed as transient, which at low~$q$ behaves as $\left(\frac{t-t_{*}}{t_{ref}}\right)^{-\frac{{\bar{\tau}_{\pi}}}{t_{ref}}}$. Note that the behavior of both modes is consistent with the earlier studies in~\cite{DeLescluze:2025jqx}, which, for the same nonthermal fixed point at $q = 0$, reported the presence of a zero mode and a tower of transient decaying modes as power-law excitations.

For the sound channel, the equation specifying the dispersion relations reads
\begin{eqnarray}
\label{eq.mastersound}
+ i \frac{\bar{\eta}}{{\cal E}} \bar{q}^2 \bar{\Omega}(\bar{q}) +\frac{1}{3} i\, {\bar{\tau}_{\pi}} \bar{q}^{2} \bar{\Omega}(\bar{q}) - i \, \bar{\tau}_{\pi}\bar{\Omega}(\bar{q})^3 \nonumber\\
+\bar{\Omega}(\bar{q})^2 - \frac{1}{3}\bar{q}^2 =0.
\end{eqnarray}
There are now three solutions, two of which correspond to sound waves
\begin{equation}
\bar{\Omega}_{\mathrm{sound}}(\bar{q}) \approx \pm \frac{1}{\sqrt{3}} \bar{q}. 
\end{equation}
Unsurprisingly, for sound waves at the leading order the behavior is $e^{\pm i\frac{1}{\sqrt{3}} q\, t}$, which, when combined with $e^{i \, q\, x^{3}}$, gives rise to the sound wave propagation familiar from equilibrium. The remaining solution of Eq.~\eqref{eq.mastersound} is another transient, which at $q = 0$ has exactly the same form as the shear. This transient is nothing else than the homogeneous isotropization we considered in Sec.~\ref{sec.isotropization}.

The last thing that we want to highlight regarding this novel instance of hydrodynamics is lurking in the rescalings~\eqref{eq.rescalingsomegaq}. As one sees, the large time limit does not commute with the small spatial momentum limit. As a result, as time progresses, at fixed $q$ higher order terms become gradually more relevant and the full form of the dispersion relation becomes important. Since our derivation is based on a truncation of the kinetic theory description, ideally one should consider the dispersion relations in the full kinetic theory. On the other hand, from the point of view of cold atomic gases experiments that have a finite lifetime, as long as one has a sufficient spatial volume at disposal, one can make the momentum of the inhomogeneity small enough so that the breakdown is not observed. On the other hand, the breakdown we predict is a robust consequence of the shear viscosity and relaxation time increasing as the system evolves, so it is an interesting phenomenon on its own.

\section{Application to QCD kinetic theory}
\label{sec.qcd}

We want to test the robustness of our derivation of far from equilibrium hydrodynamics around nonthermal fixed points against ab initio simulations. The most satisfactory solution would be to break homogeneity around a nonthermal fixed point and compare the results of an ab initio simulation with the predictions of Eq.~\eqref{eq.eom_shear}. This is numerically costly.

Instead, we will proceed with a different, qualitative route that at the same time indicates our results are more broadly applicable. To this end, we will study isotropic nonthermal fixed points in QCD kinetic theory that also give rise to the scaling~\eqref{eq.expts_UVrel}, but incorporate both binary and one-to-two collision processes of gluons. Because of the latter, the results of the previous section strictly speaking do not apply. Nevertheless, we will see that, at a qualitative level, our predictions are consistent with the QCD kinetic theory. In practical terms, we will start from an overoccupied initial condition
$
f(t_i,\vp) = n_0/g^2  \exp\left[-\vp^2/Q^2 \right]
$, with $t_i$ being the initial time of the simulation, $g^2$ the square of the coupling and $n_0$ the initial occupation. We consider $n_0=1$ and $g^2=10^{-8}$ and use the code adopted earlier in, e.g.,~\cite{Heller:2023mah}. Results will be given in units of the characteristic energy scale $Q$ as given by the maximum of $|\vp|^2 f(t_i,\vp)$.

\begin{figure*}[t!]
    \centering
    \includegraphics[width=1\linewidth]{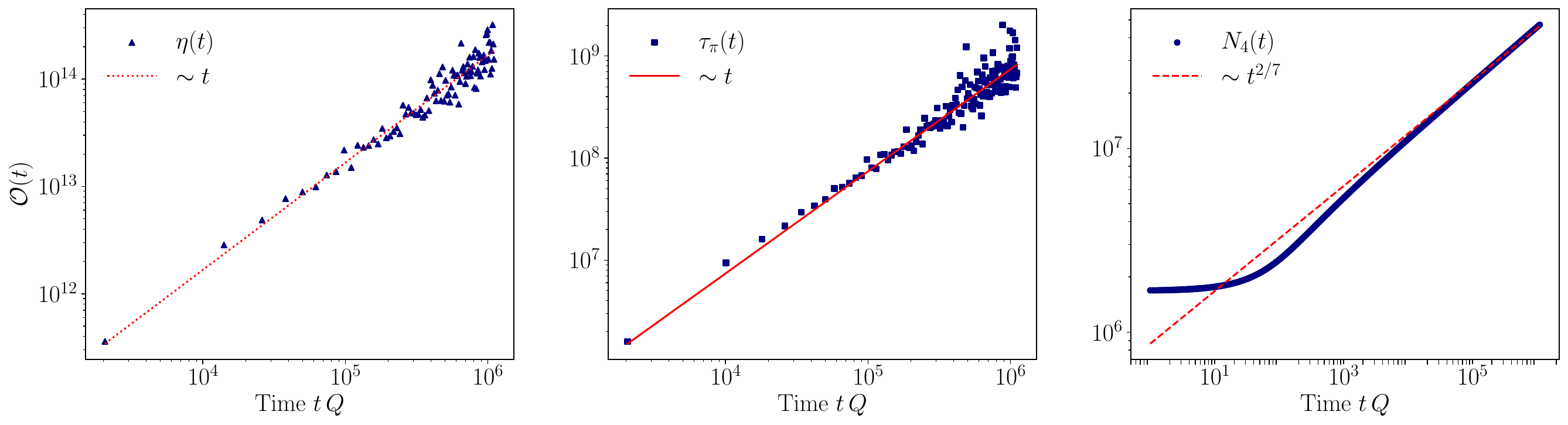}   
    \caption{Time evolution of shear viscosity (\emph{Left}), relaxation time (\emph{Center}), and $N_4 (t)$ (\emph{Right}) 
 extracted from a simulation with $g^2=10^{-8},N_c=3,n_0=1$. 
    $\delta \mathcal{C}$ is evaluated using $ 2\times 10^9$ MC samples at every 100th time step ($Q dt=0.1$, or every $tQ=10$ steps), and then averaged over 40 neighboring points to further reduce Monte Carlo fluctuations.
    }
    \label{fig.obs_rescaling}
\end{figure*}
The first test we ran is concerned with the scaling of different contributions to the expressions~\eqref{eq.transport_coeff}. Fig.~\ref{fig.obs_rescaling} shows strong indications that, given a scaling function of QCD, the transport coefficients as well as ingredients in their calculation scale appropriately. We note that, while the evolution involves also inelastic processes, we here compute only the linearized elastic collision kernel $\delta \mathcal{C}(t)$ defined in Eq.~\eqref{eq.coll_linearized} with the gluon-gluon transition matrix element of QCD kinetic theory, provided in Appendix \ref{subsec.app_relUVcascade}. The evaluation of the collision kernel $\delta \mathcal{C}$ uses Monte Carlo sampling, which leads to a spread in the uncertainty at later times. We checked that this spread can be systematically improved by increasing the number of Monte Carlo samples used in the evaluation of $\delta \mathcal{C}$. The transport coefficients are significantly affected by this spread since $\delta \mathcal{C}\sim 1/t^{5/7}$, see Eq.~\eqref{eq.linearized_collision_asymptotic_QCD} in the Appendix, which enters the expressions for the transport coefficients in the denominator. To better capture the overall scaling evolution, we average $\delta \mathcal{C}$ over 40 neighboring points. We note that such an averaging procedure is not required for calculating $ N_4(t)$ as it follows from a simple momentum integral of the distribution function. We then compare to the asymptotic fixed point scaling, where we neglect the non-universal prescaling effects including $\sim t_*$ and extract the amplitude from the bulk observables at an arbitrary reference point $\sim \mathcal{O}(t_{\mathrm{ref}})$. Overall, we observe that the scaling of different bulk observables is accurately captured by Eq.~\eqref{eq.scaling_qcd}.

\begin{figure}[h!]
    \centering
    \includegraphics[width=1\linewidth]{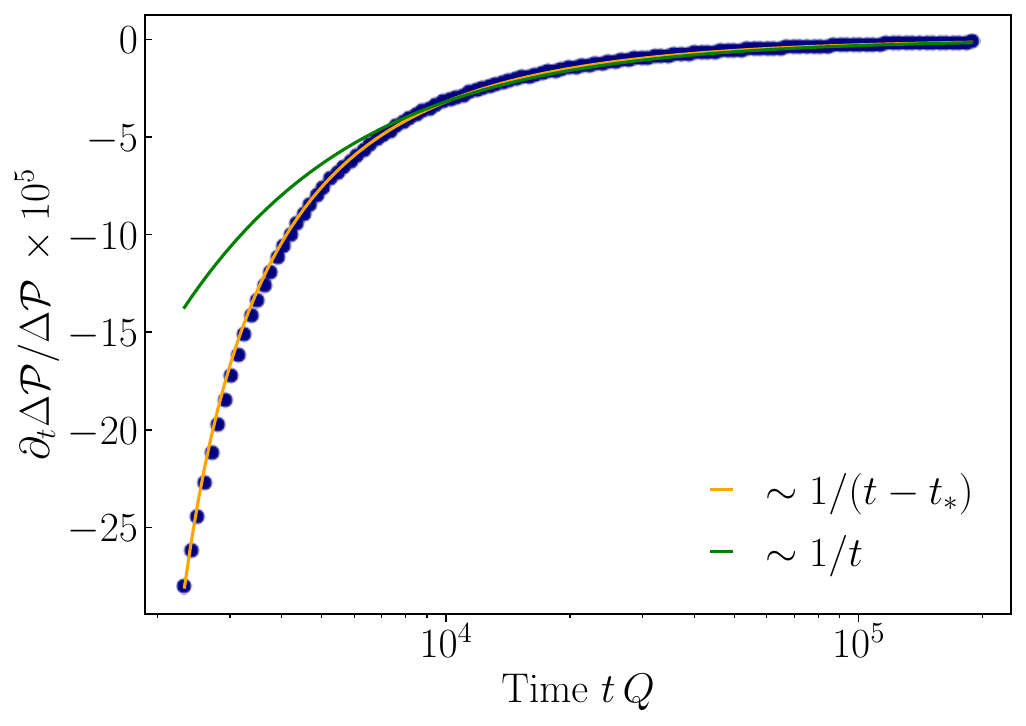}
    \caption{Evolution of the pressure anisotropy after an anisotropic perturbation has been applied to the isotropic gluonic distribution function in the vicinity of the direct energy cascade.  We compare with fixed-point scaling (green) and prescaling (orange), where the amplitude and finite-time scale $t_*\, Q\simeq 1.3 \times 10^3$ are obtained at an arbitrary reference time.}
    \label{fig.lr}
\end{figure}
The second ab initio result we obtained concerns homogeneous isotropization, discussed in Sec.~\ref{sec.isotropization}. To this end, we add an anisotropic perturbation to the gluon distribution function in the vicinity of the scaling regime:
\begin{align}
   \delta f_\vp &=  f_0(t_0,\vp)\,\frac{A}{3} \left[-(p^1)^2-(p^2)^2 + 2 (p^3)^2 \right],
\end{align}
and we follow its relaxation as witnessed by the pressure anisotropy. We apply the perturbation at a time $t_0\, Q=2\times 10^3$, when the distribution function and bulk observables, such as those in Fig.~\ref{fig.obs_rescaling}, show self-similar scaling evolution. The amplitude is set to $A=12$, though we verified that the results are independent of variations in $A$. The resulting pressure anisotropy evolution is compared to the power-law behavior of Eq.~\eqref{eq.deltaPsol} in Fig.~\ref{fig.lr}. Similar to the evolution of bulk quantities in Fig.~\ref{fig.obs_rescaling}, the late-time behavior follows fixed-point scaling $\sim 1/t$. When finite-time corrections are included via the non-universal scale $t_*$, the pressure anisotropy is accurately described even for times close to the perturbation.

\section{The nonrelativistic direct energy cascade}
The phenomenology of the direct energy cascade, discussed for relativistic systems in Sec.~\ref{sec.relativistic}, can also be found at lower energies in non-relativistic cold atom systems. Such nonthermal attractors were recently realized in cold atom platforms quenched far from equilibrium~\cite{Glidden:2020qmu,Martirosyan:2023mml,Gazo:2023exc,morenoarmijos2024}. While this attractor is also driven by a direct energy transport to smaller distance scales, the underlying quasi-particles are free non-relativistic waves with $z=2$, i.e., $\omega_\vk \sim |\mathbf{k}|^2$. 

The self-similar dynamics are governed by (perturbative) non-relativistic elastic four-wave interactions, with collision kernel~\cite{PineiroOrioli:2015cpb} 
\begin{align}
\label{eq.nonrel_L1}
    \mathcal{C}^{2\leftrightarrow 2}_{\mathrm{non-rel}}[f] &= \frac{1}{2} \int \frac{d^d \vk^\prime}{(2\pi)^d} \frac{d^d \vp}{(2\pi)^d} \frac{d^d \vp^\prime }{(2\pi)^d} W^{\mathrm{non-rel}}_{\vk \vk^\prime \leftrightarrow \vp \vp^\prime;\mathrm{U}(1)}
    \nonumber\\
    &\times   \left[(f_{\mathbf{p}}f_{\mathbf{p}^{\prime }} 
 \left( f_{\mathbf{k}}+f_{%
\mathbf{k}^{\prime }}\right) -f_{\mathbf{k}}f_{\mathbf{k}^{\prime }}\left(
f_{\mathbf{p}}+f_{\mathbf{p}^{\prime }}\right)\right] , 
\end{align}
and scattering matrix element
\begin{align}
\label{eq.nonrel_W2-2}
    W^{\mathrm{non-rel}}_{\vk \vk^\prime \leftrightarrow \vp \vp^\prime;\mathrm{U}(1)} &=4g^2 (2\pi)^{d+1}\delta^{(1)}(\omega_\vk+\omega_{\vk^\prime}-\omega_\vp-\omega_{\vp^\prime}) \nonumber\\
    &\times \delta^{(d)}(\vk+\vk^\prime-\vp-\vp^\prime),
\end{align}
where $g$ is the non-relativistic quartic coupling. A scaling analysis of this collision kernel reveals the scaling exponents
\begin{equation}
    \alpha = (d+z)\beta,\qquad \beta =-\frac{1}{3z}=-\frac{1}{6},
\end{equation}
with $d=3$ and $z=2$ for us.

For general spatial dimension $d$, this implies the following scaling behavior of the transport coefficients 
\begin{subequations} 
\label{eq.scaling_nonrel_uv}
\begin{align}
    \eta(t) &= B^{-6}(t) \bar{\eta} = \left( \frac{t-t_*}{t_{\mathrm{ref}}}\right) \bar{\eta},\label{eq.eta_scaling_nonrelativistic} \\
    \tau_\pi(t) &= B^{-6}(t)\bar{\tau}_\pi = \left( \frac{t-t_*}{t_{\mathrm{ref}}}\right)\bar{\tau}_\pi.
\end{align}
\end{subequations}
 Although the $\beta$ and $z$ exponents disagree between the relativistic and non-relativistic direct energy cascade, the transport coefficients are interestingly observed to exhibit the same scaling: they grow linearly with time.  Consequently, the time evolution of the low-energy hydrodynamic excitations in the non-relativistic energy cascade aligns with those found for the relativistic case in Sec.~\ref{subsec.modes_rel}.

Since the non-relativistic energy cascade has already been realized experimentally in well-controlled cold atom platforms~\cite{Glidden:2020qmu,Martirosyan:2023mml,Gazo:2023exc,morenoarmijos2024}, this opens up the possibility of studying the here proposed hydrodynamic signatures near nonthermal attractors. In particular, the evolution of the pressure anisotropy, as described in Eq.~\eqref{eq.deltaPsol} together with Eq.~\eqref{eq.scaling_nonrel_uv}, can be explored following an anisotropic perturbation in the atom or spin density. The time evolution of the shear viscosity, with $\eta(t)\sim t$ from Eq.~\eqref{eq.eta_scaling_nonrelativistic}, is also promising, as shear viscosity is commonly measured in cold atom systems near equilibrium~\cite{Schafer:2009dj,Adams:2012th}.

\section{Conclusions}

In the present work, we proposed that long wavelength perturbations of nonthermal fixed points are hydrodynamic in nature. This insight provides a novel extension of fluid mechanics to phenomena that are intrinsically far from equilibrium in the sense that hydrodynamic relaxation leads not to a local equilibrium, but to a local nonthermal fixed point.

We envision that bridging nonthermal fixed points and hydrodynamics will allow one to adopt ideas and techniques from the latter, venerable field into a genuinely new context of far from equilibrium evolutions. For example, the breakdown of the derivative expansion with time we encountered is reminiscent of the phenomena associated with the radius of convergence of relativistic hydrodynamics in momentum space discussed in~\cite{Withers:2018srf,Grozdanov:2019kge,Grozdanov:2019uhi}. It would be very interesting to understand this connection more thoroughly.

Our studies open new avenues for experimental studies of nonthermal fixed points in cold atom experiments, as they predict intricate dynamics associated with local conservation laws when spatial homogeneity is broken. 

The results we presented call also for new theoretical developments and analyses. The most pressing one would be going beyond the 14-moment approximation and understanding the hydrodynamic dispersion relations in full kinetic theory models. This can be achieved, e.g., by incorporating momentum dependence in the studies of~\cite{DeLescluze:2025jqx}. 

Furthermore, it would be very interesting to make a qualitative analysis of the applicability of far from equilibrium hydrodynamics using ab initio simulations with broken homogeneity. Here a good point of departure could be ideas and implementations underlying the K{\o}MP{\o}ST framework~\cite{Keegan:2016cpi,Kurkela:2018wud,Kurkela:2018vqr} (see also~\cite{Carrington:2024utf}), originally devised to deal with transverse inhomogeneity in nuclear collisions treated as a small perturbation of highly symmetric far from equilibrium boost-invariant dynamics~\cite{Bjorken:1982qr}.

Finally and in the same context, nuclear collision dynamics gives rise to an anisotropic nonthermal fixed point studied in~\cite{Baier:2000sb,Berges:2013eia,Berges:2013fga,Mazeliauskas:2018yef}. It would be very interesting to understand if hydrodynamic perturbations around this nonthermal fixed point can be framed in the language of anisotropic hydrodynamics (aHydro) that was born to deal precisely with such situations~\cite{Florkowski:2010cf,Martinez:2010sc,Alqahtani:2017mhy}.

\begin{acknowledgments}
We thank Kirill Boguslavski, Aleksi Kurkela, Matisse De Lescluze, and Aleksas Mazeliauskas for useful discussions and comments on the draft. The authors acknowledge support by the state of Baden-Württemberg through bwHPC and the German Research Foundation (DFG) through grant no 
INST 40/575-1 FUGG (JUSTUS 2 cluster), the DFG under the Collaborative Research Center SFB 1225 ISOQUANT (Project-ID 273811115) and the Heidelberg STRUCTURES Excellence Cluster under Germany's Excellence Strategy EXC2181/1-390900948. This project has received funding from the European Research Council (ERC) under the European Union’s Horizon 2020 research and innovation programme (grant number: 101089093 / project acronym: High-TheQ). Views and opinions expressed are however those of the authors only and do not necessarily reflect those of the European Union or the European Research Council. Neither the European Union nor the granting authority can be held responsible for them. G.S.D.~acknowledges support from CNPq and Funda\c c\~ao Carlos Chagas Filho de Amparo \`a Pesquisa do Estado do Rio de Janeiro (FAPERJ), grant No. E-26/202.747/2018. We would like to thank for hospitality KITP during the program ``The Many Faces of Relativistic Fluid Dynamics" supported by the National Science Foundation under Grant No. NSF PHY-1748958, where this work was initiated, and INFN Galileo Galilei Institute for Theoretical Physics during the program ``Foundations and Applications of Relativistic Hydrodynamics", where it was presented and completed.
\end{acknowledgments}

\bibliography{master.bib}

\begin{appendix}
\widetext 
\renewcommand{\theequation}{A\arabic{equation}}
\renewcommand{\thefigure}{A\arabic{figure}}
\setcounter{equation}{0}
\setcounter{figure}{0}
\makeatletter
\setlength{\parindent}{0pt}

\section{Notation}
\label{Appendix:Notation}

We define the double symmetric traceless projection of a tensor as
\begin{equation}
A^{\langle \mu \nu \rangle }\equiv \Delta ^{\mu \nu \alpha \beta} A_{\alpha \beta},
\end{equation}
with
\begin{equation}
\Delta ^{\mu \nu \alpha \beta} = \frac{1}{2}(\Delta^{\mu \alpha} \Delta^{\nu \beta} + \Delta^{\nu \alpha} \Delta^{\mu \beta}  - \frac{2}{3}\Delta^{\mu \nu} \Delta^{\alpha \beta}).
\end{equation}
We further define the projection tensor onto the $d$-space orthogonal to the four-velocity,
\begin{equation}
\Delta^{\mu \nu} = g^{\mu\nu} - u^\mu u^\nu,
\end{equation}
in terms of fluid four-velocity $u^{\mu}$ and the Minkowski spacetime metric $g_{\mu \nu}$. Finally, we denote the comoving time derivative as
\begin{equation}
\dot{A} \equiv u^{\mu}\partial_{\mu} A \equiv \frac{d}{d\tau} A,   
\end{equation}
where $\tau$ is the proper time along a flow line.

\renewcommand{\theequation}{B\arabic{equation}}
\renewcommand{\thefigure}{B\arabic{figure}}
\setcounter{equation}{0}
\setcounter{figure}{0}
\section{Derivation of Hydrodynamic equations around a nonthermal fixed point}
\label{Appendix:hydro}
In this Appendix, we discuss the derivation of the hydrodynamic equations of the type~\eqref{eq.eom_shear_thermal} around a nonthermal fixed point. The starting point is the relativistic Boltzmann equation,
\begin{equation}
k^{\mu }\partial _{\mu }f_\vk=\mathcal{C}[f],
\end{equation}%
with $k^{\mu }=(\omega _{\mathbf{k}},\mathbf{k})$ being the on-shell
four-momentum  and $%
\mathcal{C}[f]$ the collision kernel.

The shear stress tensor is determined as the following moment of $f_\vk$,
\begin{subequations}
\begin{eqnarray}
\pi ^{\mu \nu } &=&\int dKk^{\langle \mu }k^{\nu \rangle }f_\vk,\label{eq.pimunukinApp}
\end{eqnarray}
where $dK=d^{3}\mathbf{k}/[(2\pi )^{3}\omega _{\mathbf{k}}]$ is the
Lorentz-invariant phase volume. We pursue the procedure outlined in~\cite{Denicol:2010xn,Denicol:2021wod}
and obtain an equation of motion for $\pi ^{\mu \nu }$ by directly taking the
comoving derivative of~\eqref{eq.pimunukinApp}, 
\end{subequations}
\begin{equation}
\dot{\pi}^{\langle \mu \nu \rangle }=\Delta _{\alpha \beta }^{\mu \nu }\frac{d}{d\tau }%
\int dKk^{\langle \alpha }k^{\beta \rangle }f_\vk,
\end{equation}%
and using the Boltzmann equation in the form 
\begin{equation}
u^{\mu }\partial _{\mu }f_{\mathbf{k}}=-\frac{1}{u_\alpha k^\alpha}%
k^{\mu}\nabla _{\mu }f_{\mathbf{k}}+\frac{1}{u_\alpha k^\alpha}\mathcal{C}[f].
\end{equation}%
This leads to the exact equation of motion,%
\begin{align}
\label{eq.eom_pi_general}
\dot{\pi}^{\langle \mu \nu \rangle }& =\mathcal{C}_{-1}^{\mu \nu }+\frac{8}{%
15}\mathcal{E}\sigma ^{\mu \nu }-2\pi ^{\lambda \langle \mu }\omega
_{\lambda }^{\nu \rangle }-\frac{10}{7}\sigma _{\lambda }^{\langle \mu }\pi
^{\nu \rangle \lambda }  \notag \\
& -\frac{4}{3}\pi ^{\mu \nu }\theta -\rho _{-2}^{\mu \nu \alpha \beta
}\sigma _{\alpha \beta }-\Delta _{\mu _{1}\mu _{2}}^{\mu \nu }\nabla _{\mu
_{3}}\rho _{-1}^{\mu _{1}\mu _{2}\mu _{3}},
\end{align}%
where we defined the space-like gradient $\nabla _{\mu }\equiv \Delta _{\mu }^{\nu }\partial _{\nu }$, the expansion rate as the four-divergence of the four-velocity $\theta=\nabla_\mu u^\mu$, and the fluid vorticity tensor $\omega^{\mu \nu} = \left( \nabla^\mu u^\nu -\nabla^\nu u^\mu\right)/2$. Explicit details for the derivation can be found in \cite{Denicol:2010xn}.  We neglect the contributions related to heat flow, since we mostly focus on the physics of shear viscosity. Eq.~\eqref{eq.eom_pi_general} is not closed since
it depends on the collision integral $\mathcal{C}_{-1}^{\mu \nu }=\int
dKk^{\langle \mu }k^{\nu \rangle }\mathcal{C}[f]/(u_\mu k^ \mu)$ as well as on the non-hydrodynamic fields $\rho _{-1}^{\mu \nu \alpha }=\int dKk^{\langle
\mu }k^{\nu }k^{\alpha \rangle }f_{\mathbf{k}}/(u_\mu k^ \mu)$ and $%
\rho _{-2}^{\mu \nu \alpha \beta }=\int dKk^{\langle \mu }k^{\nu }k^{\alpha
}k^{\beta \rangle }f_{\mathbf{k}}/(u_\mu k^ \mu)^{2}$. The brackets denote the symmetric, traceless projection of the tensors orthogonal to the four-velocity as defined in Ref.~\cite{Denicol:2012cn}. One can close
this equation by approximating these quantities, which is traditionally
achieved using the 14-moment approximation~\cite{israel1979transient}. 

As stated in the main text of the paper, such an approach usually commences by expanding the distribution function about some background state, 
\begin{equation}
f_\vk=f_{0\vk}+\delta f_\vk,  \label{eq.lr_bg_plus_pert}
\end{equation}%
and then deriving an approximate expression for the correction $\delta f_\vk
$. In the 14-moment approximation, the correction $\delta f_\vk$ is
expanded in momentum space to second order in tensors of the particle four-momentum $k^{\mu }$, i.e., $1,k^{\mu },k^{\mu }k^{\nu }$, and truncated at
second order, 
\begin{equation}
\delta f_{\mathbf{k}} \simeq f_{0%
\mathbf{k}}\left( \epsilon +k^{\mu }\epsilon _{\mu }+k^{\mu }k^{\nu
}\epsilon _{\mu \nu }\right) ,
\end{equation}%
leading to fourteen unknown expansion coefficients, namely $\epsilon
,\epsilon _{\mu }$, and $\epsilon _{\mu \nu }$. Without loss of generality, $%
\epsilon _{\mu \nu }$ is assumed to be traceless $\epsilon _{\mu }^{\mu }=0$
(can be incorporated in $\epsilon $) and symmetric. In the presence of a
particle or net-charge current, $N^{\mu }$, the conserved currents $N^{\mu }$
and $T^{\mu \nu }$ contain a total of 14 independent degrees of freedom that
can be directly matched to the expansion coefficients. The scalar
coefficient $\epsilon $ and the vector coefficient $\epsilon ^{\mu }$ can be
directly related to the bulk viscous pressure and heat flow dissipative
contributions, respectively \cite{israel1979transient,deGroot:80relativistic,Denicol:482569}, and, thus, will not be relevant in this
analysis. On the other hand, the tensor coefficient $\epsilon ^{\mu \nu }$
is directly matched to the shear stress tensor of the fluid, $\pi ^{\mu \nu }
$, and contains the dominant contribution to the hydrodynamic equations.
This matching procedure can be solved directly following the procedure
outlined in \cite{Denicol:2012cn,Denicol:2021wod,denicol2022microscopic}, which can be applied without \textit{any} modifications even if the background distribution $f_{0\mathbf{k}}$ is not the local equilibrium distribution function -- the fundamental assumption employed in this method is that $f_{0\mathbf{k}}$ must be isotropic. Thus, for any isotropic $f_{0\mathbf{%
k}}$, one can demonstrate that 
\begin{equation}
\label{eq.phi_14momentsApp}
    \delta f_\vk = f_{0\vk}\phi_\vk = f_{0\vk} \frac{2p^\mu p^\nu}{15 N_4(t)} \pi_{\mu \nu}(t),
\end{equation}
with $N_4(t)=\int dK |\vk|^4 f_{0\vk}$. 
We note that the inclusion of additional terms in the moment expansion $\sim k^{\mu }k^{\nu
}k^{\rho },\dots $ will increase the precision, but their contribution is
suppressed in thermal hydrodynamics as small momentum scales dominate the
physics due to the underlying separation of scales.

Considering this more general application of the method of moments, one can generalize the derivation of fluid dynamics proposed by Israel and Stewart to a more general class of systems that are not necessarily close to equilibrium. In this work, we perform this task for systems that are in the vicinity of a nonthermal attractor, which arises in weakly-coupled systems that are
over-occupied, $f_{\mathbf{k}}\gg 1$. The dynamics are usually
dominated by elastic two-to-two collisions, which can be captured by a
collision term of the form~\cite{PineiroOrioli:2015cpb}
\begin{equation}
C\left[ f\right] =\frac{1}{2}\int dK^{\prime }dPdP^{\prime }W_{\mathbf{kk}%
\prime \rightarrow \mathbf{pp}\prime }I[f](\mathbf{k},\mathbf{k}^{\prime },%
\mathbf{p},\mathbf{p}^{\prime })\;,  \label{Col_term}
\end{equation}%
where $W_{\mathbf{kk}\prime \rightarrow \mathbf{pp}\prime }$ is the
Lorentz-invariant transition rate. For an
over-occupied system, the gain-loss contribution can  be approximated by its classical-wave limit 
\begin{align}
& I[f](\mathbf{k},\mathbf{k}^{\prime },\mathbf{p},\mathbf{p}^{\prime })  =f_{\mathbf{p}}f_{\mathbf{p}^{\prime }}[1+f_{\mathbf{k}}][1+f_{\mathbf{k}%
^{\prime }}]-f_{\mathbf{k}}f_{\mathbf{k}^{\prime }}[1+f_{\mathbf{p}}][1+f_{%
\mathbf{p}^{\prime }}] \\
& \approx f_{\mathbf{p}}f_{\mathbf{p}^{\prime }}\left( f_{\mathbf{k}}+f_{%
\mathbf{k}^{\prime }}\right) -f_{\mathbf{k}}f_{\mathbf{k}^{\prime }}\left(
f_{\mathbf{p}}+f_{\mathbf{p}^{\prime }}\right) .
\end{align}%
As stated in the main text, we focus on nonthermal attractors associated with the
turbulent transport of mode energy to smaller distance scales through a
direct energy cascade, described by
weakly-interacting relativistic massless scalars with a
transition element in the $O(N)$ model~\cite%
{PineiroOrioli:2015cpb}, see Eq.~\eqref{eq.transition_matrix} in the main text of the paper.

In a next step, we plug the expansion of Eq.~\eqref{eq.lr_bg_plus_pert} together with the explicit perturbation of Eq.~\eqref{eq.phi_14momentsApp}
into the right-hand side of Eq.~\eqref{eq.eom_pi_general}, which will allow us to close the equation as the higher-order collision kernel and hydrodynamic fields will be expressed in terms of $\pi^{\mu \nu}$. The collision kernel contribution $\mathcal{C}^{\mu \nu}_{-1}$ linearized in the perturbation is then expressed in terms of $\pi^{\mu \nu}$ as
\begin{equation}
    \mathcal{C}^{\mu \nu}_{-1} = - \delta C \frac{15}{2N_4(t)}\pi^{\mu \nu},
\end{equation}
with the linearized collision kernel being defined as
    \begin{align}
    \label{eq.coll_linearized}
    \delta \mathcal{C} &=\int dK dK^\prime dP dP^\prime  \frac{1}{u_\mu k^\mu} k_{\langle \alpha} k_{\beta \rangle} W_{\vk \vk^\prime \to \vp \vp^\prime} \left[ -\frac{1}{10}f_{0\vp} f_{0\vp^\prime} f_{0\vk} \left(k^{\langle \alpha} k^{\beta \rangle} + 2p^{\langle \alpha } p^{\beta \rangle}\right)\right.\nonumber\\
    &\left. -\frac{1}{10} f_{0\vp} f_{0\vp^\prime} f_{0\vk^\prime} \left( k^{\prime \langle \alpha} k^{\prime \beta\rangle} + 2p^{\langle \alpha} p^{\beta \rangle}\right) +\frac{1}{5} f_{0\vk} f_{0\vk^\prime} f_{0\vp} \left(k^{\langle \alpha} k^{\beta \rangle} + k^{\prime \langle \alpha} k^{\prime \beta\rangle} +p^{\langle \alpha} p^{\beta \rangle} \right) \right].
\end{align}
We note that since $\mathcal{C}^{\mu \nu}_{-1}$ is a symmetric, traceless rank-2 tensor orthogonal to the four-velocity field, it must vanish when integrated with any isotropic momentum distribution function, even though the collision term itself does not vanish in such a non-equilibrium state. Thus, the leading contribution to this collision integral will indeed be the dissipative contribution shown above, that has the same structure of the collision integral calculated around local equilibrium. Nevertheless, the coefficient $\sim \delta \mathcal{C}$ contains information about the nonthermal attractor of the background state and acts as a generalized transport coefficient. 

The evolution equation of the hydrodynamic shear stress follows then directly from Eq.~\eqref{eq.eom_pi_general}
\begin{align}
\label{eq.eom_shearApp}
    \tau_\pi(t) \dot{\pi}^{\langle \mu \nu\rangle} &=- \pi^{\mu \nu} +2\eta(t) \sigma^{\mu \nu}+\tau_\pi(t) \left[-\frac{4}{3}\pi^{\mu \nu}\theta-\frac{10}{7} \pi^{\lambda \langle \mu}\sigma^{\nu\rangle}_\lambda -2\pi^{\lambda\langle \mu} \omega^{\nu\rangle}_\lambda \right],
\end{align}
where we identified the shear viscosity and relaxation time transport coefficient by direct comparison with the thermal case~\eqref{eq.eom_shear_thermal},
\begin{subequations}
\label{eq.transport_coeff_App}
\begin{align}
    \eta(t) &= \frac{8}{15^2} \frac{\mathcal{E} N_4(t)}{\delta \mathcal{C}(t)},\\
        \tau_\pi(t) &= \frac{2}{15}\frac{N_4(t)}{\delta \mathcal{C}(t)}.
\end{align}
\end{subequations}

\renewcommand{\theequation}{C\arabic{equation}}
\renewcommand{\thefigure}{C\arabic{figure}}
\setcounter{equation}{0}
\setcounter{figure}{0}
\section{Scaling analysis}
In this Appendix, we analyse the scaling of bulk observables under the scaling ansatz~\eqref{eq.prescaling}. We use the notation $\bar{\vp}=B(t)\vp$ and the scaling relation $A(t)=B^\sigma(t)$ with $\sigma=d+z$ for energy number conservation. 

The energy and fourth moment of the distribution function follow as
\begin{subequations}
\begin{align}
    \mathcal{E}(t)&=\int \frac{d^3\vp}{(2\pi)^3} \omega_\vp f(t,\vp) =A(t) B^{-(d+z)}(t) \int \frac{d^3 \bar{\vp}}{(2\pi)^3} \omega_{\bar{\vp}} f_S(\bar{\vp})\equiv B^{\sigma-(d+z)}(t) \bar{\mathcal{E}},\\
    N_4(t) &= \int \frac{d^3 \vp}{(2\pi)^3 \omega_\vp} |\vp|^4 f(t,\vp)  =A(t) B^{-(d+4-z)}(t) \int \frac{d^3 \bar{\vp}}{(2\pi)^3 \omega_{\bar{\vp}}} |\bar{\vp}|^{4} f_S(\bar{\vp}) \equiv B^{\sigma-(d+4-z)}(t) \bar{N}_4 ,
\end{align}
\end{subequations}
in the vicinity of the fixed point. To derive the scaling of transport coefficients from Eq.~\eqref{eq.transport_coeff}, we need to analyze the time evolution of the linearized collision kernel~\eqref{eq.coll_linearized}. We will discuss this scaling analysis for the relativistic and non-relativistic energy cascade in turn.

\subsection{Relativistic energy cascade}
\label{subsec.app_relUVcascade}
We consider first the relativistic energy cascade with the application of QCD kinetic theory in Sec.~\ref{sec.qcd} in mind. Starting from Eq.~\eqref{eq.coll_linearized}, we evaluate the expressions as
\begin{align}
    \delta \mathcal{C}^{\mathrm{rel,UV}}(t)
    &=\int \frac{d^3 \vk}{(2\pi)^3 \omega_\vk} \frac{d^3 \vk^\prime}{(2\pi)^3 \omega_{\vk^\prime}} \frac{d^3 \vp}{(2\pi)^3 \omega_\vp} \frac{d^3 \vp^\prime}{(2\pi)^3 \omega_{\vp^\prime}} \frac{1}{\omega_\vk} \vk^2 W_{\vk \vk^\prime \to \vp \vp^\prime} \left[ -\frac{1}{10}f_{0\vp} f_{0\vp^\prime} f_{0\vk} \left(\vk^2 \frac{2}{3} +  2 \vp^2 (\mu^2_{\vk \vp}-1/3)\right)\right.\nonumber\\
    &\left. -\frac{1}{10} f_{0\vp} f_{0\vp^\prime} f_{0\vk^\prime} \left( \vk^{\prime 2}(\mu^2_{\vk \vk^\prime}-1/3) + 2 \vp^2 (\mu^2_{\vk \vp}-1/3)\right) \right.\nonumber\\
    &\left. +\frac{1}{5} f_{0\vk} f_{0\vk^\prime} f_{0\vp} \left(\frac{2}{3} \vk^2 + \vk^{\prime 2} (\mu^2_{\vk,\vk^\prime}-1/3) +\vp^2 (\mu^2_{\vk \vp}-1/3) \right) \right],
\end{align}
where we used the tensor contractions
\begin{align}
    k_{\langle \alpha} k_{\beta \rangle} p^{\langle \alpha} p^{\beta \rangle} &= \Delta_{\alpha \beta \mu \nu } k^\mu k^\nu \Delta^{\alpha \beta \rho \sigma} p_\rho p_\sigma = \Delta^{\rho \sigma}_{\mu \nu} k^\mu k^\nu p_\rho p_\sigma \\
    &=\frac{1}{2} \left[2\left(k\cdot p-\omega_\vk \omega_\vp \right)^2-\frac{2}{3} \left(k^2-\omega^2_\vk\right)\left(p^2-\omega^2_\vp\right) \right]\\
    &= \vk^2 \vp^2 \left( \mu^2_{\vk \vp} -\frac{1}{3}\right),
\end{align}
together with 
\begin{subequations} 
\begin{align}
k^2-\omega^2_\vk &= \omega^2_\vk-\vk^2 -\omega^2_\vk = -\vk^2, \\
k\cdot p -\omega_\vk \omega_\vp &=-\vk \cdot \vp = - |\vk| |\vp| \mu_{\vk \vp}
    k_{\langle \alpha } k_{\beta \rangle} k^{\langle \alpha} k^{\beta \rangle} 
    \frac{2}{3} \vk^4,\\
    k_{\langle \alpha} k_{\beta \rangle} k^{\prime \langle \alpha} k^{\prime \beta \rangle} &= \vk^2 {\vk^\prime}^2 \Big(\mu^2_{\vk,\vk^\prime}-\frac{1}{3}\Big),
\end{align}
\end{subequations}
$\mu_{\vk \vp}\equiv \cos(\theta_{\vk,\vp})$, and $\theta \sphericalangle (\vk,\vp)$. The transition matrix element for the energy cascade is generally (for the relativistic and the non-relativistic energy cascade) given by
\begin{equation}
    W_{\vk \vk^\prime \to \vp\vp^\prime} =|\mathcal{M}|^2_{\vk \vk^\prime \to \vp \vp^\prime} (2\pi)^{d+1} \delta^{(1)}(\omega_{\vk} + \omega_{\vk^\prime}-\omega_\vp -\omega_{\vp^\prime}) \delta^{(d)}(\vk+\vk^\prime-\vp-\vp^\prime), 
\end{equation}
with theory-specific scattering matrix element $|\mathcal{M}|^2$. For the relativistic $O(N)$ model and for QCD, we have
\begin{subequations}
    \begin{align}
        |\mathcal{M}_{\mathrm{O}(N)}|^2  &=
        \lambda^2 \frac{N+2}{6N^2}, \\
        |\mathcal{M}^{gg}_{gg}|^2(k,k^\prime,p,p^\prime) &=16 (N^2_c-1) N^2_c g^4 \left( 3-\frac{s_M t_M}{u^2_M} - \frac{s_M u_M}{t^2_M} - \frac{t_M u_M}{s^2_M}\right),
    \end{align}
\end{subequations}
where $N_c$ denotes the number of colors and $u_M,s_M,t_M$ the usual Mandelstam variables. Here we only give the vacuum expression for the $2\leftrightarrow 2$ gluon scattering matrix element averaged over spin and color degrees of freedom. In practice, $|\mathcal{M}^{gg}_{gg}|$ has an infrared divergence which is regulated by the medium Debye mass. We will neglect the corresponding sub-leading logarithmic correction, see~\cite{Brewer:2022vkq,Heller:2023mah}, for the purposes of the scaling analysis. The leading order QCD collision kernel receives, besides $\mathcal{C}^{2\leftrightarrow 2}$, a contribution from effective $1\leftrightarrow 2$ scatterings, which we will also drop in the scaling analysis as it scales like the $2\leftrightarrow 2$ kernel, see~\cite{Heller:2023mah}. Parametrizing the scaling behavior of the collision kernel as
\begin{equation}
    \mathcal{C}[f](t,\vp) = A(t)^{\mu_\alpha} B(t)^{\mu_\beta} \mathcal{C}[f_S](\bar{\vp})=B(t)^{\sigma \mu_\alpha+\mu_\beta}\mathcal{C}[f_S](\bar{\vp}),
\end{equation}
where $1/\beta = (1-\mu_\alpha)\sigma -\mu_\beta$. For both collision kernels, we identify $\mu_\alpha=3$ and $\mu_\beta=-2d+5z$. The scaling of the linearized collision kernel contribution for the relativistic energy cascade can then be written as
\begin{align}
    \delta \mathcal{C}^{\mathrm{rel,UV}}(t) &\equiv B^{3\sigma -4d+5z-4+(d+z)} \delta \bar{\mathcal{C}}^{\mathrm{rel,UV}}(t) =B^{-1/\beta+4z-6}(t) \delta \bar{\mathcal{C}}_{-1,\mathrm{UV}}\\
    &=B^{5}(t) \delta \bar{\mathcal{C}}^{\mathrm{rel,UV}}\\
    &= \left( \frac{t-t_*}{t_{\mathrm{ref}}}\right)^{-\frac{5}{7}} \delta \bar{C}^{\mathrm{rel,UV}}.\label{eq.linearized_collision_asymptotic_QCD}
\end{align}
Altogether, using Eq.~\eqref{eq.transport_coeff}, this implies for the transport coefficients that
\begin{align}
    \eta^{\mathrm{rel,UV}}(t) &=\frac{8}{15^2} \frac{B^{\sigma-(d+z)}\bar{\mathcal{E}} B^{\sigma-(d+4-z)}(t)\overline{N}_4}{B^{9z-4}(t) \delta \bar{\mathcal{C}}^{\mathrm{rel,UV}}(t)}\equiv B^{-7z}(t) \bar{\eta}^{\mathrm{rel,UV}}\\
    &=B^{2(1-z)+1/\beta}(t) \bar{\eta}^{\mathrm{rel,UV}}\\
    \tau^{\mathrm{rel,UV}}_{\pi}(t) &= \frac{2 }{15} \frac{B^{\sigma-(d+4-z)}(t)\overline{N}^4}{B^{9z-4}(t) \delta \bar{\mathcal{C}}^{\mathrm{rel,UV}}(t)} \equiv B^{-7z}(t) \bar{\tau}^{\mathrm{rel,UV}}_{\pi}\\
    &=B^{2(1-z)+1/\beta}(t) \bar{\tau}^{\mathrm{rel,UV}}_\pi.
\end{align}
This reproduces the scaling in Eq.~\eqref{eq.scaling_qcd} for $z=1$ and $\beta=-1/7$.

\subsection{Non-relativistic energy cascade}
\label{app.nonrelativistic}
The kinetic description of a non-relativistic system differs to that of a relativistic one in the appearance of $\omega_\vk$ factors in the Boltzmann equation and in the momentum integration measures, see Eq.~\eqref{eq.nonrel_L1} and Eq.~\eqref{eq.nonrel_W2-2}.
We can then obtain the non-relativistic energy cascade scaling from the corresponding relativistic expressions by dropping the $\omega_\vk$ factors
\begin{subequations}
    \begin{align}
    N_4(t) &= \int \frac{d^3 \vp}{(2\pi)^3} \vp^4 f(t,\vp) = A(t) B^{-(d+4)}(t) \int \frac{d^3 \bar{\vp}}{(2\pi)^3} \bar{p}^{4} f_S(\bar{\vp}) \equiv B^{\sigma-(d+4)}(t) \overline{N}_4,\\
        \delta \mathcal{C}^{\mathrm{non-rel,UV}}(t) &=B^{-4d+3\sigma-4+d+z}(t)\delta \bar{\mathcal{C}}^{\mathrm{non-rel,UV}} = B^{4(z-1)}(t) \delta \bar{\mathcal{C}}^{\mathrm{non-rel,UV}},
\end{align}
\end{subequations}
where we used that the non-relativistic collision kernel scales with $\mu_\alpha=3$ and $\mu_\beta=d+z-3d$. This implies the following scaling of transport coefficients
\begin{subequations}
    \begin{align}
        \eta^{\mathrm{non-rel,UV}}(t) &= B^{\sigma-(d+z)+\sigma-(d+4)-(4z-4)}(t)\bar{\eta}^{\mathrm{non-rel,UV}} =B^{-3z}(t) \bar{\eta}^{\mathrm{non-rel,UV}},\\
        \tau^{\mathrm{non-rel,UV}}_\pi(t) &= B^{\sigma-(d+4)-(4z-4)}(t)\bar{\tau}^{\mathrm{non-rel,UV}}_\pi =B^{-3z}(t) \bar{\tau}^{\mathrm{non-rel,UV}}_\pi,
    \end{align}
\end{subequations}
which reproduces Eq.~\eqref{eq.scaling_nonrel_uv} with $z=2$ and $\beta=-1/6$.

\end{appendix}

\end{document}